\documentclass[journal=jacsat,manuscript=article]{achemso}
\usepackage{graphicx,adjustbox,multirow,latexsym,amssymb,amsmath,bbold,paralist,braket,appendix}
\usepackage{hyperref}
\usepackage[dvipsnames]{xcolor}
\usepackage{booktabs}

\usepackage[version=3]{mhchem} 

\newcommand{\OFN}{C$_{10}$F$_{8}$}
\newcommand{\Phth}{C$_{8}$H$_{6}$N$_{2}$}
\newcommand{\muC}{$\mu$C~cm$^{-2}$}
\newcommand{\pcm}{cm$^{-1}$}
\newcommand{\Zstar}{$\bf{Z^{\star}}$}

\title{A Piezoelectric Molecular Cocrystal with Unconventional $\pi$-Stacking}

\author{Samuel G. Dunning}
\affiliation{Earth and Planets Laboratory, Carnegie Institution for Science, Washington, D.C. 20015, United States}
\email{sdunning@carnegiescience.edu}
\author{Aldo Raeliarijaona}%
\affiliation{Earth and Planets Laboratory, Carnegie Institution for Science, Washington, D.C. 20015, United States}
\author{Piotr A. Guńka}%
\affiliation{ Faculty of Chemistry, Warsaw University of Technology, 00-664 Warszawa, Poland}
\author{Anirudh Hari}%
\affiliation{Earth and Planets Laboratory, Carnegie Institution for Science, Washington, D.C. 20015, United States}
\author{Dongzhou Zhang}%
\affiliation{Hawaii Institute of Geophysics and Planetology, School of Ocean and Earth Science and Technology, University of Hawaii at Manoa, Honolulu, HI 96822, United States}
\author{R.~E.~Cohen}%
\affiliation{Earth and Planets Laboratory, Carnegie Institution for Science, Washington, D.C. 20015, United States}
\email{rcohen@carnegiescience.edu}
\author{Timothy A. Strobel}%
\affiliation{Earth and Planets Laboratory, Carnegie Institution for Science, Washington, D.C. 20015, United States}
\email{tstrobel@carnegiescience.edu}

\date{May 9, 2025}
             
\begin{document}
\begin{abstract}
We demonstrate the crystallization of a polar octafluoronaphthalene (OFN, \OFN)--phthalazine (Phth, \Phth) cocrystal, formed in a 1:2 ratio by slow evaporation. The crystal structure and vibrational properties of the cocrystal were determined using powder/single-crystal X-ray diffraction (XRD) and Fourier-Transform Infrared (FTIR) spectroscopy, and confirmed with density functional theory (DFT) and density functional perturbation theory (DFPT) calculations. The molecular $\pi$-stacking of aromatic rings is unconventional compared with other arene--perfluoroarene cocrystals. Phth molecules are offset and misaligned with respect to the major axis of OFN due to electrostatic repulsion between N and F atoms, enabling overall electric polarization attributed to the dipole moment of Phth. Our calculations show that OFN:2Phth is an insulator with a band gap of $\sim$2.4 eV. The electric polarization was calculated to be 7.1 \muC, while the shear piezoelectric coefficient ($d_{34}$) may be as large as 11.4 pC N$^{-1}$.  
\end{abstract}

\maketitle


\section{\label{sec:level1}Introduction\protect \lowercase{}}

Materials that crystallize in certain non-centrosymmetric space groups are of great technological interest due to their inherent spontaneous polarization which gives rise to unusual electronic and non-linear optical properties. Notably, these systems respond to temperature (pyroelectricity), mechanical stress, and electric fields (piezoelectricity), making them ideal candidates for transducers, frequency multipliers and a range of other applications.\cite{Kumar_2021,Kishore_2021,Shiroshi_2021,Gupta_2021,Meena_2014,Ortega_2015}

Many piezoelectrics that find use in devices are inorganic oxides (e.g., perovskites) that typically contain toxic and/or rare metals\cite{Cohen_1992, Nuraje_2013}. Molecular organic piezoelectrics are attracting growing interest as a safer alternative due, in part, to their reduced toxicity and relative abundance of component elements. Other advantages include their low densities, mechanical flexibility, and ease of synthesis.\cite{Stroppa2011,Horiuchi_2008,Pan2024}

However, several barriers remain in the search for organic piezoelectric crystals that are competitive with their inorganic analogues. Most notably, the absence of reliable synthetic strategies to produce organic piezoelectrics remains a significant challenge. Although it is relatively simple to synthesize polar organic molecules, these systems tend to crystallize in centrosymmetric structures where dipole--dipole interactions cancel out. Therefore, reliable synthetic strategies are essential to control the formation of organic piezoelectrics and design new systems for advanced applications. 

Currently, the most reliable method of organic piezoelectric synthesis is the formation of charge-transfer cocrystals. These systems are comprised of electron-rich and electron-poor aromatic compounds which, when combined, form 1D stacks of alternating aromatic ring systems, organized through electron donor/acceptor interactions. \cite{Liu_2022,Horiuchi_2008} These cocrystals undergo a neutral to ionic transition (e.g., TTF-CA, M$_{2}$P-DMTCNQ),\cite{Oison2003,Giovannetti2009,Ferrari2022,Horiuchi2022} wherein intermolecular electron transfer triggers piezoelectric behavior. Charge-transfer cocrystals represent an important class of materials, however, they have certain limitations. Notably, charge-transfer cocrystals are typically formed from organic molecules that are themselves non-polar. Accordingly, the piezoelectric behavior of these systems is not derived from the inherent polarity of constituent molecules, but is rather induced through formal charge transfer between molecules or structural rearrangements.  These systems rely on a narrow charge gap between donor and acceptor molecules to induce the transition, leading to low breakdown fields and high losses. 

An alternative crystal engineering strategy for the development of new organic polar materials is illustrated by the formation of arene:perfluoroarene cocrystals. These systems---which self-assemble into ordered, alternating molecular stacks---can be formed cleanly and efficiently through the slow evaporation of a mixed solution of molecular species, or even through grinding.\cite{SAUNDERS2013162} While similar in concept to charge-transfer systems, in that both systems combine electron-rich and electron-deficient aromatic molecules, arene:perfluoroarene cocrystals have a large HOMO--LUMO gap that inhibits charge transfer.\cite{Mal2019} Arene:perfluoroarene systems can thus incorporate a wide variety of molecular constituents compared to charge-transfer complexes, which rely heavily on the use of electron donor/acceptor molecules such as tetracyanoquinodimethane (TNCQ) and tetrathiafulvalene (TTF) to form cocrystals with HOMO--LUMO gaps suitable for charge transfer.

Recent reports have shown that arene:perfluoroarene cocrystals can also exhibit piezoelectricity and ferroelectric transitions  via mechanisms other than the neutral-to-ionic transitions in charge-transfer cocrystals. For example, a cocrystal of acenaphthene and 2,3,5,6-tetrafluoro-7,7,8,8-tetracyanoquinodimethane was found to exhibit switchable polarization at room temperature ($T_{C}$= 68 $^\circ$C), facilitated by temperature-induced rotation of acenaphthene molecules.\cite{Wiscons_2018}  As with most molecular compounds, arene:perfluoroarene cocrystals  typically crystallize in centrosymmetric space groups.\cite{Marder2002,Marder2001} However, there have been a handful of reported polar arene:perfluoroarene cocrystals  \cite{Morimoto_2008,Batsanov_2001,Wu_2021,MAKAROV_2006,COLLINGS200237}. Interestingly, a small subset of these systems incorporate molecular constituents that possess a permanent dipole.\cite{Morimoto_2008}

Herein, we describe the formation of a new polar arene:perfluoroarene cocrystal derived from phthalazine (Phth, \Phth) and octafluoronaphthalene (OFN, \OFN).  This OFN:2Phth system, which can be synthesized cleanly at scale, exhibits a calculated permanent polarization of 7 \muC, and a piezoelectric coefficient that can be as large as 11.4 pC N$^{-1}$. Arene:perfluoroarene cocrystallization thus represents a new approach for molecular ferroelectrics that allows for the incorporation of both polar and non-polar organic molecules and is amenable to crystal engineering methods.

\section{Methods}
\subsection{Experimental Methods}
Octafluoronaphthalene (OFN, $>$98\%, TCI America) and Phthalazine (Phth, $>$98\%, TCI America) were used as received without further purification. Like many organic molecules, Phth and OFN crystallize in the centrosymmetric space groups $Pbca$ and $P2_{1}/c$, respectively.\cite{Donohue:a24253} However, when combined and recrystallized via slow evaporation from solution in acetone, new cocrystals precipitate as colorless plates. Equimolar amounts of OFN (54 mg, 0.2 mmol) and Phth (26 mg, 0.2 mmol) were dissolved in 10 mL of acetone. The container was then covered with foil, and a needle was used to create several small holes in the covering. The samples were left to crystallize via slow evaporation, producing small plate-like crystals after several days. Curiously, when the precursors are combined in a 1:1 molar ratio, the resulting cocrystals form with the composition 1OFN:2Phth. If the starting OFN:Phth ratio is varied prior to crystallization, the resulting product contain only OFN:2Phth together with whichever precursor was in excess. 

As with molecular OFN and Phth, OFN:2Phth cocrystals readily sublime under ambient conditions, which presents challenges for structural and properties characterization. We, therefore, rapidly collected single-crystal diffraction data using a synchrotron source. A suitable single crystal was selected under the microscope and mounted on a Mitigen MicroMesh$^{TM}$ using a thin layer of mineral oil. X-ray diffraction data were collected on beamline 13-BM-C ($\lambda$ = 0.4340 \AA) using a Pilatus3 1M detector. Two omega scans with different detector positions (2$\theta = 0^\circ$ and 2$\theta =20^\circ$) were obtained between $\omega = -$150$^\circ$ to $+$150$^\circ$ in 0.5$^\circ$ steps with an exposure time of 1 s per frame. The frames were then imported into CrysAlisPRO \cite{noauthor_crysalispro_2023}. Data reduction was carried out independently on the two scans and the resulting $hkl$ files were scaled and merged using SORTAV invoked from WinGX \cite{blessing_data_1987,blessing_dreadd_1989,farrugia_wingx_2012}. Crystal structure was solved using SHELXT 2018/2 via intrinsic phasing and refined using SHELXL 2018/3 via least-squares refinement \cite{sheldrick_shelxt_2015,sheldrick_crystal_2015}. Both programs were invoked from Olex2 ver. 1.5 \cite{dolomanov_olex2_2009}. CCDC 2429146 contains supplementary crystallographic data for this paper.\cite{ccdc} 

Powder X-ray diffraction data were collected on crushed crystals using a Bruker D8 Discover microdiffraction system comprised of a Cu $K_\alpha$ microfocus source with a 2-mm beam collimator and a Vantec500 area detector. Obtained 2D diffraction images were combined and integrated using the Bruker DIFFRAC.EVA software package.\cite{eva}

Fourier-Transform infrared spectroscopy measurements were obtained using a Bruker Vertex
spectrometer with Hyperion microscope and MCT detector between 500--8,300 cm$^{-1}$ with a resolution of 4 cm$^{-1}$. 

Thermogravimetric analysis (TGA) and differential scanning calorimetry (DSC) measurements were carried out using a TA Instruments SDT-650 TGA/DSC. Experiments were carried out under inert Ar atmosphere with a flow rate of 100 mL/min. Samples were heated between 30--350 $^\circ$C after equilibration at 30 $^\circ$C with a ramp rate of 10 $^\circ$C/min.

Tests for second-harmonic generation by frequency doubling were conducted by placing a cocrystal within a sample chamber comprised of two diamond windows, and an IPG single-mode ytterbium fiber laser (1070 nm) operating at $\sim$30 W was focused onto the sample using a long-working-distance infrared objective while the sample was imaged for the emission of visible light.

\subsection{Theoretical Methods}

Density functional theory calculations were performed using Quantum Espresso ({\sc QE})\cite{QE-2009,QE-2017,QE-exa} and the Vienna ab initio simulation package ({\sc VASP})\cite{VASPI,VASPII,VASPIII}. Optimal norm-conserving Vanderbilt (ONCV)\cite{ONCVI,ONCVII} pseudopotentials, and the projector augmented-wave method (PAW)\cite{PAW}  were used for {\sc QE} and {\sc VASP}, respectively. The Brillouin zone (BZ) was sampled using a 6$\times$4$\times$4 Monkhorst-Pack grid\cite{MP_1976} and the plane-wave expansion was truncated using a cut-off energy of E$_{\text{cutoff}}$ = 950 eV. We chose hard pseudopotentials to avoid any overlaps between the PAW's. We used the non-local first-principles van der Waals DF2 functional\cite{Lee_2010} self-consistently for both {\sc QE} and {\sc VASP}. The atomic positions were relaxed with forces were lower than 2.5 meV/\AA. 

The polarization was calculated within the Modern theory of polarization\cite{King-Smith1993,Resta1994} (Berry phase calculations). The infrared (IR), Raman frequencies, and the dynamical Born effective charges were calculated using the density functional perturbation theory (DFPT, i.e. linear response)\cite{Baroni_2001} implemented in the PHonon ({\sc PH}) package of {\sc QE}. The piezoelectric coefficients and the elastic moduli were computed using DFPT\cite{Wu_2005} as implemented in {\sc VASP}\cite{VASPI,VASPII,VASPIII}. 
With the strain $\eta$ and electric field $\mathcal{E}$ as perturbations, the relaxed-ion piezoelectric coefficient ($e$) and the elastic modulus ($C$) are given by
\begin{eqnarray}
    e_{\alpha\beta} = -\frac{\partial^{2}{\tilde{E}}}{\partial{\mathcal{E_{\alpha}}}\partial{\eta}_{\beta}},\\
     C_{\alpha\beta} = \frac{\partial^{2}\tilde{E}}{\partial{\eta_{\alpha}}\partial{\eta_{\beta}}}\bigg|_{\mathcal{E}},
\end{eqnarray}
where $\tilde{E}(\eta,\mathcal{E}) = \min\limits_{u}{E(u,\eta,\mathcal{E})}$, is the energy minimum with respect to the atomic displacement $u$.\cite{Wu_2005} The piezoelectric strain constant ($d$) was calculated from the piezoelectric stress constant ($e$) using the relation:
\begin{equation}
    d_{\alpha\beta} = \sum\limits_{\mu}S^{(\mathcal{E})}_{\beta\mu}e_{\alpha\mu},
\end{equation}
where $S^{(\mathcal{E})}$ is the elastic compliance tensor at fixed electric field, which is just the inverse of the elastic modulus $S^{(\mathcal{E})} = (C^{(\mathcal{E})})^{-1}$.
By definition, piezoelectricity is the response of polarization to strain at constant electric field. Calculations with strain can, however, lead to nonphysical responses. For example, the polarization can change under a uniaxial or biaxial stretch because of the change of lattice parameter even though the relative atomic positions didn't change. The physically relevant piezoelectric responses are the ones that do not include these spurious polarization changes with respect to strain. These are the {\it proper} piezoelectric coefficients\cite{Saghi-Szabo1998,Vanderbilt_2000} and they are the appropriate coefficients for comparisons with experimentally measured piezoelectric responses. The piezoelectric coefficients calculated from DFPT using {\sc VASP} and reported here are the proper piezoelectric coefficients.

\section{Results and Discussion}

We find that OFN:2Phth cocrystals form in the non-centrosymmetric space group $P2_{1}$ with cell parameters $a$ $\approx$ 7.00\AA, $b$ $\approx$ 11.15\AA, $c$ $\approx$ 14.43\AA, and $\beta$ $\approx$ 94.9$^\circ$, as determined by single-crystal diffraction. The relaxed structure of OFN:2Phth from first-principles calculations matches the experimental structure well. The $-$4.3\% difference in calculated volume is within the expected error between static first-principles calculations and the experimental measurement performed at room temperature. Experimental and calculated crystallographic parameters for OFN:2Phth are provided in \textbf{Table \ref{Table1}}.

\begin{figure}
    \centering
    \includegraphics[width=0.75\columnwidth]{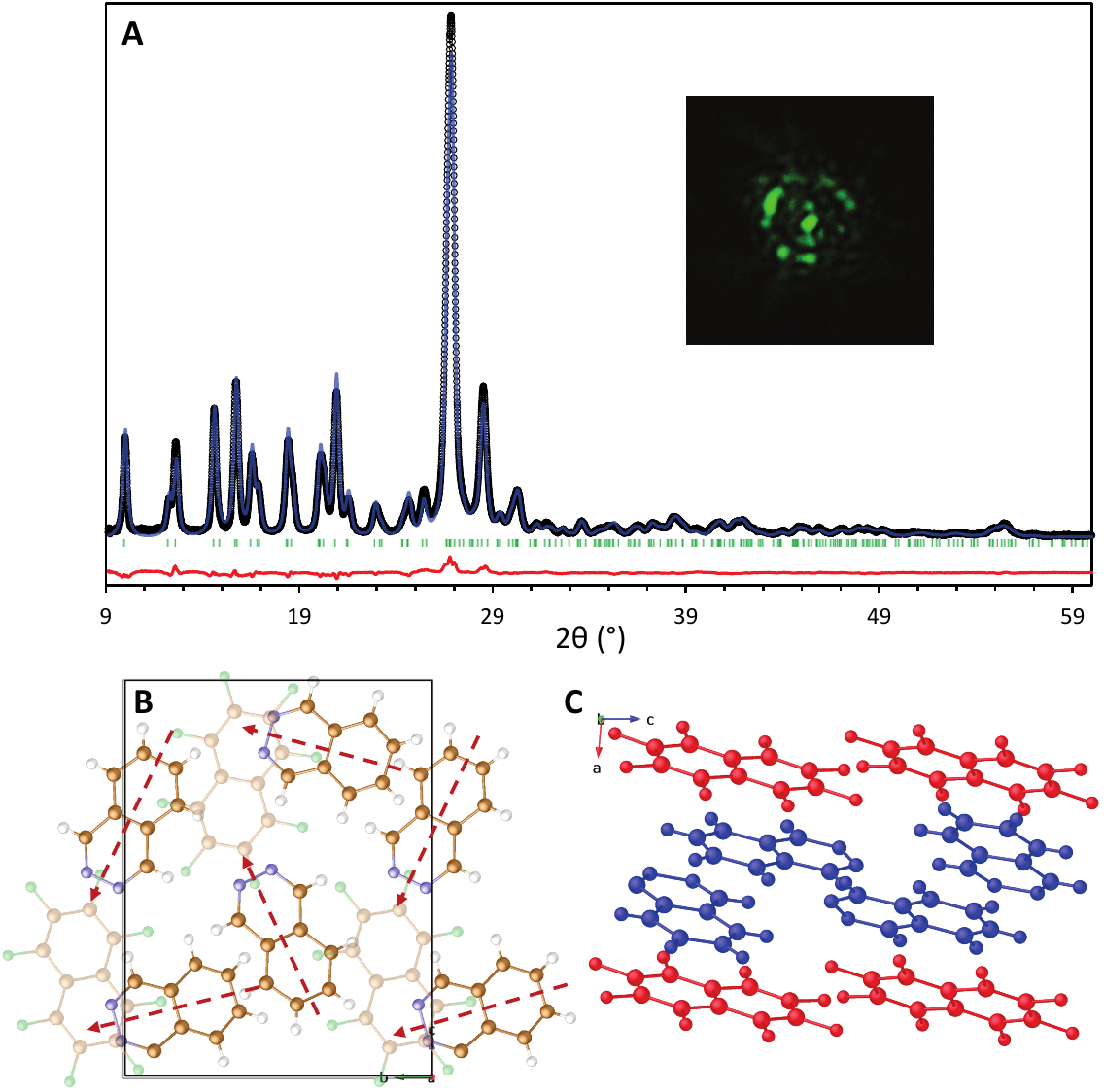}
    \caption{(\textbf{A}) Comparison of the powder X-ray diffraction pattern (Cu $K_\alpha$ radiation) of bulk OFN:2Phth (black points) with the Rietveld refinement carried out using the single-crystal structural parameters  (blue line). The residual is represented by the red trace at the bottom. Bragg reflections are indicated by green bars. The inset shows second-harmonic generation upon exposing the cocrystal to 1070 nm light. (\textbf{B}) Single-crystal structure of OFN:2Phth viewed along the crystallographic \textit{X}-axis. Red arrows indicate the direction of the molecular dipole in Phth molecules. The opacity of OFN molecules has been reduced for clarity of depth. (\textbf{C}) Alternative view of the single-crystal structure of OFN:2Phth showing the layered structure containing alternating sheets of OFN (red) and Phth (blue) molecules.}
    \label{Fig1}
\end{figure}

Rietveld refinement of powder X-ray diffraction data obtained from a sample of crushed crystals indicates that this structure is representative of the bulk material (\textbf{Fig.\ref{Fig1}A}). Upon exposure to a 1070 nm laser, the cocrystals exhibit frequency doubling and the emission of 535 nm light, confirming the polar nature of the crystal structure. Thermogravimetric analysis of the cocrystal product shows a two-step decomposition process with onset temperatures of 95 $^\circ$C and 188 $^\circ$C, assigned to the sublimation of Phth and OFN molecules, respectively (see \textbf{Supporting Information}, \textbf{Fig. S1}).  Samples also readily sublime at ambient temperature. 

\begin{table}[!ht]
\centering
\normalsize
\caption{Experimental and DFT-computed (in brackets)  crystal structure data for OFN:2Phth}
\begin{tabular}{ l  l }
\toprule
Empirical Formula &  \OFN : 2\Phth \\
CSD Number & 2429146 \\
$T$/K & 293 \\
Crystal system &  Monoclinic\\
Space group &  $P2_{1}$\\
$a$/\AA~ & 7.0063(3) ~~~[6.82]\\
$b$/\AA~ & 11.1567(4) ~~[11.16]\\
$c$/\AA~ & 14.4383(7) ~~[14.17]\\
$\beta$/$^\circ$ ~ & 94.872(4) ~~~[94.27]\\
$V$/\AA$^3$~ & 1124.53(8) ~~[1075.50]\\
$Z$ &  2\\
$\rho$ calc/g$\cdot$cm$^{-3}$ & 1.572\\
$\mu$ / mm$^{-1}$ & 0.053\\
$F$(000) &  536.0\\
Radiation & Synchrotron, $\lambda$=0.4340\\
2$\theta$ range for data collection /$^\circ$  & 1.729--17.5\\
\multirow{3}{6em}{Index Ranges} & $-$9$\leq h \leq$9\\
                                & $-$15$\leq k \leq$15\\
                                & $-$20$\leq l \leq$20\\
Reflections collected & 19762\\
Independent Reflections & 6117 ($R_{int}$=0.066)\\
Data/Restraints/parameters & 6117/1/344\\
Goodness-of-fit on F$^2$ & 0.94\\
\multirow{2}{12em}{Final $R$ indexes ($I$ $\geq$ 2$\sigma$($I$))} & $R_1$=0.047\\
                                                                 & $wR_2$=0.101\\
\multirow{2}{12em}{Final $R$ indexes (all data)} & $R_1$=0.051\\
                                                & $wR_2$=0.129\\
Largest diff. peak/hole /  e \AA $^{-3}$ & $+$0.18/$-$0.19\\
Flack parameter & $-$3.9(10)\\
\bottomrule
\end{tabular} 
\label{Table1}
\end{table}

Consistent with other arene:perfluoroarene cocrystals, OFN:2Phth crystallizes with an AB layered structure consisting of alternating sheets of OFN and Phth molecules. Most interestingly, unlike molecular phthalazine, which crystallizes as discrete dimers with offsetting molecular dipoles,\cite{Huiszoon:a09406} the Phth molecules in OFN:2Phth are arranged in such a way that produces a dipole along the crystallographic $Y$-axis (\textbf{Fig.\ref{Fig1}B}). The stacking of aromatic molecules is governed by the balance between repulsive $\pi - \pi$ and attractive $\pi - \sigma$ interactions.\cite{Hunter_Sanders_1990} Differences in $\pi - \pi$ interactions between molecules leads to different stacking arrangements of aromatic rings, such as benzene dimers where parallel, staggered, or T-shaped stacking arrangements are observed.\cite{Hunter_Sanders_1990,Lee_2007} The uniform distribution of partial charge can produce ``sandwich-like'' $\pi-\pi$ stacking as observed in OFN:naphthalene (C$_{10}$F$_{8}$:C$_{10}$H$_{8}$), a different cocrystal where the OFN and naphthalene molecules neatly stack in AB layers along the $X$-axis in a 1:1 ratio.\cite{Potenza:a12440, Ward_2019} 

While OFN:2Phth exhibits the expected AB layering pattern, such parallel,``sandwich-like'' stacking along a particular direction does not exist for polar OFN:2Phth. Each of the OFN molecules within OFN:2Phth are, in fact, oriented in stacks down the crystallographic $X$-axis (\textbf{Fig.\ref{Fig1}B)}. They are layered between offset and rotated Phth molecules with twice as many molecules per layer. This unconventional stacking of the polar cocrystal can be explained using the electrostatic argument of Hunter and Sanders as follows.\cite{Hunter_Sanders_1990} In neatly stacked cocrystals such as OFN:naphthalene, the H and F atoms (capping members of both molecules) carry opposite partial charges. The electronegativity difference between carbon and the capping members produces electron-rich (naphthalene) and electron-deficient (OFN)  $\pi$ systems. This leads to a balance between attractive and repulsive interactions and the stabilization of sandwich-like $\pi-\pi$ stacking. For the case of OFN:2Phth, both molecules incorporate regions that carry a partial negative charge. For phthalazine, this partial negative charge occurs at the diazine group, while the entire octafluoronapthalene molecule is terminated by electron-withdrawing F atoms. Interestingly, while the the lone pairs on the F atoms are  sp\textsuperscript{3} hybridized, and therefore project into the Phth layers above and below, the Phth diazine group contains  sp\textsuperscript{2} hybridized N atoms wherein the lone pair electrons extend out into the Phth plane. The relative positioning of the lone pair electrons may account for the unusual structure, as electrostatic repulsion between F and  atoms likely causes the respective molecules to misalign. Further, the diazine group in each Phth is pointed directly towards an electron deficient OFN ring in the neighboring layer. It is possible that the diazine groups can, at least partially, stabilize the electron deficient OFN rings. Each OFN unit contains two fused rings, which are then each stabilized by one Phth diazine unit, which may explain the observed 1OFN:2Phth ratio. Nevertheless, the cocrystals readily sublime under ambient conditions, indicating diminished $\pi$ interactions, consistent with the observed structure.  Alternatively, one can consider dipole interactions among the molecules, which favor either parallel or antiparallel arrangements depending sensitively on molecular distances.\cite{LinesGlass1977} Non-dipolar interactions between molecules can also help stabilize polar or antipolar arrangements, as described above. \\

\begin{figure}
    \centering
    \includegraphics[width=0.5\columnwidth]{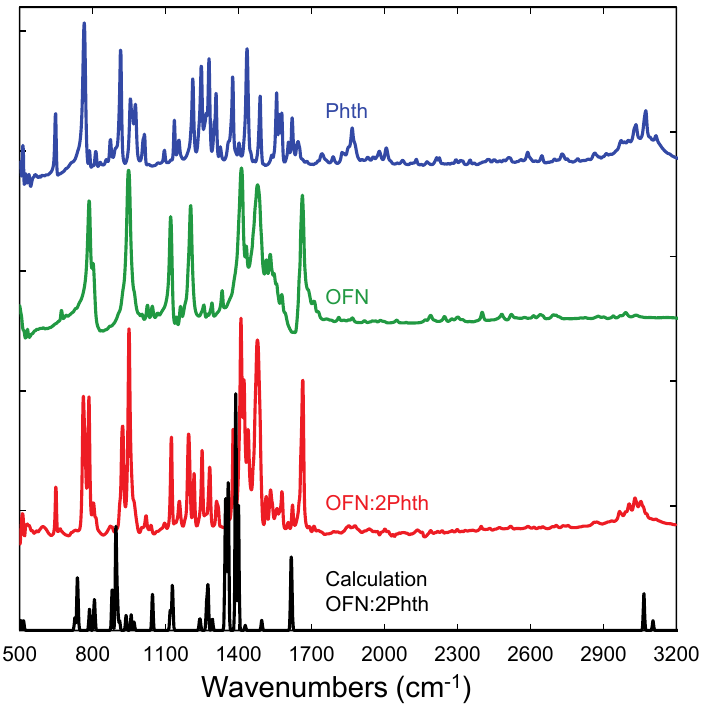}
    \caption{Experimental FTIR absorbance spectrum of OFN:2Phth cocrystal (red) compared with the calculated spectrum from DFPT (black), as well as experimental spectra from OFN (green) and Phth (blue). The calculated IR spectrum is a convolution of Gaussian peaks with an arbitrary linewidth of 2.5 \pcm. Spectra are shifted vertically for clarity.}
    \label{Fig2}
\end{figure}


The vibrational properties of OFN:2Phth were investigated using FTIR spectroscopy and DFPT calculations. The calculated IR spectrum compares favorably with the experimental spectrum (\textbf{Fig.\ref{Fig2}}). A detailed table of observed frequencies is provided in \textbf{Table S1}. Molecular contributions to the FTIR spectrum of the cocrystal are readily identified by comparison with the individual OFN and Phth starting materials (\textbf{Fig.\ref{Fig2}}). Matching the FTIR frequencies of the cocrystal to those of the individual OFN and Phth components, we find that the four cocrystal modes calculated at 513, 648, 762, and 1377 \pcm~can be attributed to Phth. The high-frequency cocrystal modes in the range of 2950--3100 \pcm~originate from the C--H vibrations of Phth molecules, and are softened by up to 60 \pcm~in the cocrystal. Additionally, vibrational modes in the cocrystal represented by peaks at 785, 949, 1124, 1490, 1477, 1512, 1531, 1579, and 1668 \pcm~are attributed to OFN modes.\cite{NIST_2024}. Atomic displacements associated with vibrations in DFPT calculations confirm FTIR mode assignments. Moderate frequency shifts between the calculated and experimental spectra arise from use of the harmonic approximation and finite temperature differences.

\begin{figure}
    \centering
    \includegraphics[width=0.5\columnwidth]{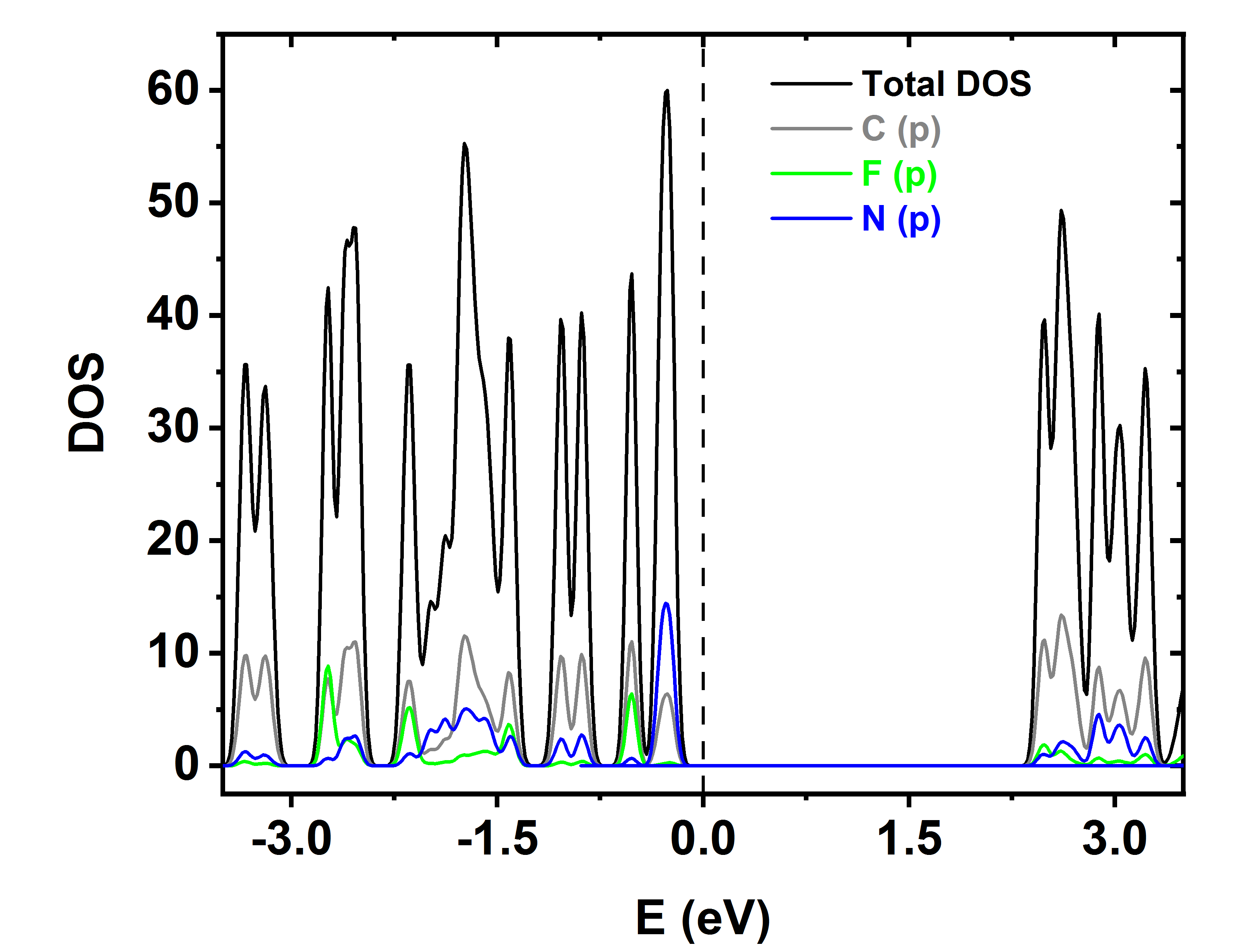}
    \caption{Computed density of states (DOS) and partial density of states (PDOS) of OFN:2Phth. The dashed line indicates the Fermi level. The gray, green and blue lines represent the PDOS of $p$ electrons belonging to C, F and N atoms, respectively. The DOS were computed using an $8\times 4 \times 4$ $k$-point grid.}
    \label{Fig3}
\end{figure}

The calculated density of states (DOS) of OFN:2Phth shows that the cocrystal is an insulator with a sizable band gap of 2.4 eV (\textbf{Fig.\ref{Fig3}}). Because DFT computations are known to underestimate the energy band gap, the true gap must be $>$2.4~eV. This is consistent with the colorless nature of the crystals. 

Given the non-centrosymmetric crystal structure of OFN:2Phth, polar properties were investigated using DFT calculations. In addition to Berry phase calculations, the electric polarization can also be calculated with the dynamical Born effective charge using

\begin{equation}
  \textbf{P} = \frac{1}{\Omega}\sum^{N}_{i}{{\bf Z}^{\star}_{i} {\bf u}_{i}},
\end{equation}
where $\Omega$ is the unit cell volume, the atomic index $i$ runs to the number of atoms $N$ in the unit cell, $\bf{Z^{\star}_{i}}$ and ${\bf u}_{i}$ are the dynamical Born effective charge and the displacement of the $i^{th}$ atom, respectively. We confirmed that the polarization calculated with \Zstar~reproduces what we computed from Berry phase calculations. 


The total polarization of the cocrystal can  be split into two parts:
\begin{equation}
    {\bf P}_{\text{tot}}={\bf P}_{\text{OFN}} + {\bf P}_{\text{Phth}},
\end{equation}
    where {\bf P}$_{\text{OFN}}$ and {\bf P}$_{\text{Phth}}$ are the calculated contributions to the polarization from the two OFN and four Phth molecules in the unit cell, respectively, using the effective charges \Zstar. Based on molecular dipoles, the total polarization likely originates from Phth molecules, and {\bf P}$_{\text{Phth}}$ is naively expected to be close to {\bf P}$_{\text{tot}}$ and {\bf P}$_{\text{OFN}}$ is expected to be negligible. However, when using Born effective charges, this is not the case. By symmetry, the non-polar $X$ and $Z$ directions have zero polarization components, so they cancel for two  ${\bf P}_{\text{OFN}}=\left[0.02, 0.28, 0.04\right]$  C~$m^{-2}$  in the crystal coordinate system, and four ${\bf P}_{\text{Phth}}=\left[-0.02, -0.21, -0.04\right]$ C~$m^{-2}$. {\bf P}$_{\text{OFN}}$ and {\bf P}$_{\text{Phth}}$ have non-equal values with opposite signs in the polar direction. The Born effective charges $\bf{Z^{\star}_{i}}$, quantify how much of the polarization changes as one moves the $i^{th}$ atom. ${\bf P}_{\text{OFN}}$ and ${\bf P}_{\text{Phth}}$ from our calculations thus show the change of polarization as one moves the OFN and the Phth molecules, respectively. Or alternatively, we see that the non-polar molecule is polarized by its motion in the crystal due to the non-polar environment.

We find several nonzero piezoelectric coefficients that are comparable to other molecular piezoelectrics (\textbf{Table \ref{Table:Piezo}}). For instance, the magnitude of the largest piezoelectric coefficient of OFN:2Phth ($d_{34}=11~\text{pC N}^{-1}$) is comparable to diisopropylammonium bromide (DIPAB) ($d_{33}=11~\text{pC N}^{-1}$)\cite{DaWeiFu_2013,You_2017}; $d_{22}$ and $d_{23}$ of the cocrystal are also comparable to that of croconic acid ($d_{33}=5~\text{pC N}^{-1}$).\cite{Horiuchi_2010} \footnote{Voigt notation is used for the piezoelectric tensor elements, with the first index being the Cartesian polarization or electric field direction, and the second index being the strain matrix elements, $1=11$, $4=23$, etc. The $e=dP/d\epsilon$ elements are the change in polarization with strain, and the $d=d\epsilon/dE$ elements are the strain with applied electric field.}

The piezoelectric response is sufficient for these simple to produce crystals to be used in piezoelectric applications. While the fact that these crystal easily sublimate means they would require encapsulation, this discovery opens up a new set of systems to study for potential applications. By varying the molecular components for these cocrystals, there is likely a great variety of potentially useful materials in this family with improved stability, polarization, electrochemical response, and mechanical properties.

\begin{table}[!ht]
\centering
\normalsize
\caption{Direct and inverse piezoelectric coefficients of OFN:2Phth cocrystals computed using DFPT as implemented in {\sc VASP}.}
\begin{tabular}{ c  c  c  c  }
\toprule
\multicolumn{2}{c}{Direct (\muC)} & \multicolumn{2}{c}{Inverse ($\text{pC N}^{-1}$)} \\
\hline
$e_{21}$ & $-$0.11  & $d_{21}$ & $-$0.31 \\
$e_{22}$ & 2.74  & $d_{22}$ &  1.55 \\
$e_{23}$ & $-$2.14 & $d_{23}$ & $-$1.25\\
$e_{14}$ & 0.15  & $d_{14}$ & 1.65 \\
$e_{16}$ & 0.11  & $d_{16}$ & 2.54 \\
$e_{25}$ & $-$0.002 & $d_{25}$ & 1.08 \\
$e_{34}$ & 2.43  & $d_{34}$ & 11.4 \\
$e_{36}$ & 0.48  & $d_{36}$ & 3.29 \\
\bottomrule
\end{tabular}
\label{Table:Piezo}
\end{table}

\section{Conclusions}
Using slow evaporation of a solution of polar phthalazine and nonpolar octafluoronaphthalene molecules, a novel non-centrosymmetric cocrystal with the formula OFN:2Phth was produced. This cocrystal forms in the polar crystallographic space group $P$2$_{1}$ and exhibits unconventional $\pi$-stacking interactions due to electrostatic repulsion between nitrogen and fluorine, and vibrational modes can be largely understood through contributions from the component molecules.  The polar cocrystal has a calculated DFT band gap of 2.4~eV and an electric polarization of $7~$\muC. OFN:2Phth also exhibits a large shear piezoelectric coefficient $d_{34}$ of $11~$pC/N. The cocrystallization of arene:perfluoroarene molecules represents a novel route to synthesize organic piezoelectrics.

\vspace{3mm}

\begin{acknowledgement}
This work was supported as part of the Carnegie Summer Undergraduate Internship (SURI) program. The authors acknowledge funding support from The
Arnold and Mabel Beckman Foundation (Arnold O. Beckman
Postdoctoral Fellowship in Chemical Sciences Program).  A.R. and R.C. acknowledge support from the U. S. Office of Naval Research Grant N00014-20-1-2699. Computations were supported by high-performance computer time and resources from the DoD High Performance Computing Modernization Program (HPCMP).

Portions of this work were performed at GeoSoilEnviroCARS (The University of Chicago, Sector 13), Advanced Photon Source (APS), Argonne National Laboratory. GeoSoilEnviroCARS was supported by the National Science Foundation--Earth Sciences (EAR--1634415). This research used resources of the Advanced Photon Source, a U.S. Department of Energy (DOE) Office of Science User Facility operated for the DOE Office of Science by Argonne National Laboratory under Contract No. DE-AC02-06CH11357.

\end{acknowledgement}
\bibstyle{achemso}
\bibliography{Main}

\end{document}


\maketitle



\pagebreak

\begin{table}[h!]
\centering
\caption{Experimental IR frequencies of OFN:2Phth cocrystal, OFN, and Phth.}
\begin{tabular}{| c | c | c | }
\hline
 Cocrystal (cm$^{-1}$) &  OFN (cm$^{-1}$) &  Phth (cm$^{-1}$) \\
 \hline
 512 &         &  513 \\
 532 & 532  &         \\
 601 & 590  &         \\
 647 &         &  646 \\
 761 &         &  766 \\
 784 & 787  &         \\
 804 &         &  809 \\
 871 &         &  874 \\
 923 &         &  914 \\
 948 &  949 &         \\
1016 &         & 1012 \\
1020 & 1024 &         \\
1039 & 1045 &         \\
1095 & 1093 &         \\
1124 & 1120 & 1136 \\
1155 &         & 1153 \\
1193 &         & 1212 \\
1196 & 1201 &         \\
1250 &         & 1248 \\
1280 &         & 1278 \\
1309 &         & 1307 \\
1376 &         & 1376 \\
1410 &         & 1410 \\
1439 &         & 1435 \\
1477 & 1477 & 1489 \\
1512 & 1513 &         \\
1531 & 1527 &         \\
1579 & 1579 &         \\
1668 & 1662 &         \\
2968 &         & 2970 \\
3031 &         & 3002 \\
3033 &         & 3072 \\
3056 &         & 3116 \\
\hline
\end{tabular}
\label{TableS1}
\end{table}

\vspace{20mm}

\begin{figure}
    \centering
    \includegraphics[width=\columnwidth]{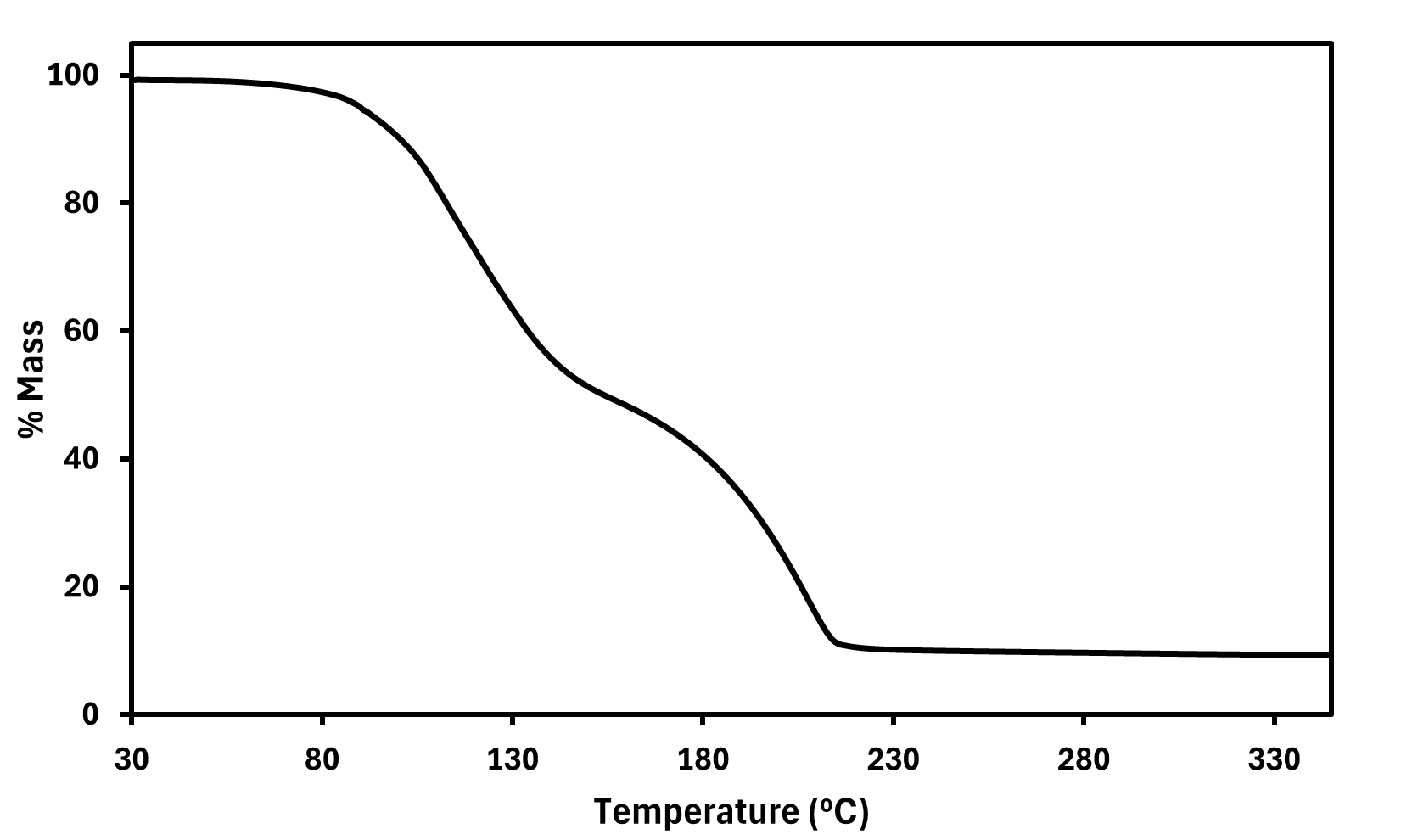}
    \caption{Thermogravimetric analysis of OFN:2Phth collected between 30--350 $^\circ$C under an inert Ar atmosphere. The two-step decomposition is indicative of the loss of OFN and Phth respectively}
    \label{SFig1}
\end{figure}